\documentclass[11pt]{article}

\begin{document}

%\markboth{Edmundo M. Monte}
%{Topological and Geometrical Properties of Brane-worlds}

%\catchline{}{}{}{}{}
%
%%%%%%%%%%%%%%%%%%%%%%%%%%%%%%%%%%%%%%%%%%%%%%%%%%%%%%%%%%%%%%%%%%%%
%\title{TOPOLOGICAL AND GEOMETRICAL PROPERTIES OF BRANE-WORLDS}

%\author{EDMUNDO M. MONTE}
%\footnote{edmundo@fisica.ufpb.br and edmundomonte@pq.cnpq.br}
%\address{Departamento de Fisica, Universidade Federal da Paraiba, Cidade Universitaria \\
%Joao Pessoa, Paraiba, 58059-970, Brasil.\\
%\email{edmundo@fisica.ufpb.br\footnote{another e-mail: edmundomonte@pq.cnpq.br and monteedmundo@gmail.com}} }
%\maketitle
\title{TOPOLOGICAL AND GEOMETRICAL PROPERTIES OF BRANE-WORLDS}
\author{EDMUNDO M. MONTE\thanks{e-mail: edmundo@fisica.ufpb.br}\\
Departamento de F\'{\i}sica, Universidade \\Federal da
Para\'{\i}ba, 58059-970, Jo\~{a}o Pessoa, Para\'{\i}ba, Brasil}

\maketitle
%\begin{history}
%\received{(Day Month Year)}
%\revised{(Day Month Year)}
%\end{history}

\begin{abstract}

We study the geometrical and topological properties of the bulk (environment space) when we
modify the geometry or topology of a brane-world. Through the
characterization of a spherically symmetric space-time as a local brane-world immersed
into six dimensional pseudo-Euclidean spaces, with different signatures of the bulk, 
we investigate the existence of a topological difference in the
immersed brane-world. In particular the Schwarzschild's brane-world  and its
Kruskal (or Fronsdal) brane-world extension are  examined from point of view of the immersion formalism. 
We prove that there is a change of signature of the bulk when we consider a local isometric 
immersion and different topologies of a brane-world in that bulk. 
 
\end{abstract}

%\keywords{Immersion of space-times; brane-world; Schwarzschild's space-time.}

\section{Introduction}

The immersion problem of space-times in spaces with higher dimension 
has been required in subjects linked to
minimal class of the immersion, extrinsic gravity and theory of
strings. Nowadays the immersion problem emerged with
brane-world theory - a theory that lead us the idea of
unification of fundamental interactions using extra dimensions.
Such a model has been showing positive in the sense that we find
perspectives and probably deep modifications in the physics, such
as: unification in a TeV scale, quantum gravity in this scale and
deviation of Newton's law of gravity for small distances. 

A brane-world may be regarded as a space-time locally immersed in a
higher dimensional space, the bulk,  solution of higher
dimensional Einstein's equations. Furthermore, the immersed
geometry  is assumed to  exhibit quantum  fluctuations with
respect to  the extra dimensions at the TeV scale of  energies.
Finally, all  gauge interactions belonging to the standard model
must remain confined to the four-dimensional space-time.
Contrasting with  other  higher dimensional theories, the extra
dimensions may be large and even infinite, with the possibility of
being observed by  TeV accelerators. The integrability conditions of immersion
relate the  bulk  geometry  to the brane-world geometry. \cite{MonteIJGMMP,Eis}

On this context it is interesting to study the geometric and
topological properties of the bulk (environment space) when we
modify the geometry or topology of the brane-world. \cite{Kanti}
Specifically, for example: we consider a brane-world immersed into bulk, if
we modify the topology of the brane-world, what will happen with
the signature of the bulk's metric?

In the present paper we show that if we
have $Y:(M^{n},g)\longrightarrow (\bar{M}^{D},\bar{g})$ a local
isometric immersion and a topology $\tau'_{\eta}$ of $Y(W^{n})$,
($W^{n}$ a neighborhood of $p \in M^{n}$), different of the
induced topology $\tau_{\eta}$ of the $Y(W^{n})$ and the
determinants of the tensor metric $g_{ij}$ and ${g'_{\mu \nu}}$, (a coordinates transformation on $g_{ij}$), 
are not equal in sign at a point, then there is a change of
signature of the bulk's metric. Furthermore it is made an application that
results from an example of general relativity in brane-worlds
context. On the example we use the Schwarzschild space-time as a
brane-world and we apply a change of topology via extension of
Kruskal (or Fronsdal) metric obtaining in this form signature change of the
six-dimensional flat bulk, where is immersed this brane-world.

\section{The Schwarzschild's Extension from Immersion Formalism}

We are going to return to the following problem:
{\em Determine the
physical space outside of an approximately spherical body with
mass $M $}. The physical space is  modeled through
a 4-dimensional space-time,  solution of Einstein equations, whose geometry
is  described with good approximation by 
Schwarzschild's  solution,  representing the  empty
space-time with spherical symmetry  outside of a body with spherical
mass, where  $ M =c^{2}mG^{-1}$, $c$ is the speed of light and  $G$ is the
gravitational constant. \cite{HE}

We know that in spherical coordinates $(t,r,\theta ,\phi )$, the regions
 $r=0$ and $r=2m$ are singular. When we remove the surface $r=2m$, the manifold becomes separated in two
disconnected components, one for $2m<r<\infty $ and the other for $0<r<2m $.
Since we  are  dealing with the existence  of the  metric
associated to a  physical space, we require a connected space. Therefore, we define the following regions:\cite{Oneill}\\
a)The exterior Schwarzschild space-time  $(V_{4},g)$:\\
$ V_{4}=P_{I}^{2}\times S^{2}\;$; $\; P_{I}^{2}=\{(t,r)\in \Re^{2}|\; r>2m\}\;$\\
b) The Schwarzschild black hole  $(B_{4},g)$:\\
$ B_{4}=P_{II}^{2}\times S^{2}\;$, 
$\; P_{II}^{2}=\{(t,r)\in \Re^{2} |\; 0<r<2m\}\;$\\
In both cases, $S^{2}$  is the sphere of radius $r$ and  the metric $g$ is
given by the usual metric of Schwarzschild. We know that $(B_{4},g)$ and $(V_{4},g)$ may be
extensible for $r=2m$. The extension of $(V_{4},g)$ was calculated by Kruskal but it was suggested by C. Fronsdal one year before. 
\cite{Kruskal,Fronsdal} 

Now we use the isometric immersion formalism to establish the extension of
$(E,g)=([P_{I}^{2}\cup P_{II}^{2}]\times S^{2},g)$, denoted  by
$(E',g')=(Q^{2}\times S^{2},g')$, where  $Q^{2}$  is the Kruskal plane.

Consider two known
isometric immersions of space-time $(E,g)$ into a pseudo Euclidean 
manifold
of six dimensions, with different signatures:\\
%\vspace{2mm}
- The Kasner immersion: \cite{Kasner}
\[
ds^{2}=\;
dY_{1}^{2}+dY_{2}^{2}-dY_{3}^{2}-dY_{4}^{2}-dY_{5}^{2}-dY_{6}^{2}.
\]
-The Fronsdal immersion: \cite{Fronsdal}
\[
ds^{2}=\;{d{Y'}_{1}}^{2}-{d{Y'}_{2}}^{2}-{d{Y'}_{3}}^{2}-
{d{Y'}_{4}}^{2}-{d{Y'}_{5}}^{2}-{d{Y'}_{6}}^{2}, 
\]
Respectively given by (using  $2m \equiv 1$)
\begin{equation}
\left\{
\begin{array}{l}
Y_{1}=(1-1/r)^{1/2}\mbox{cos}t\\
\vspace{1mm} Y_{2}=(1-1/r)^{1/2}\mbox{sin} t\\
\vspace{1mm}Y_{3}=f(r),
\;\;
(df/dr)^{2}=\frac{1+4r^{3}}{4r^{3}(r-1)}\\
\vspace{1mm}
Y_{4}=r\mbox{sin}\theta \mbox{sin}\phi\\
\vspace{1mm}
Y_{5}=r\mbox{sin}\theta \mbox{cos}\phi\\
\vspace{1mm}Y_{6}=r\mbox{cos}
\theta\vspace{1mm}\end{array}\right.\;\;\;\;
\mbox{and}\;\;\;\left\{
\begin{array}{l}
Y'_{1}=2(1-1/r)^{1/2}\mbox{sinh}(t/2)\\
\vspace{1mm}
Y'_{2}=2(1-1/r)^{1/2}\mbox{cos}
h(t/2)\\
\vspace{1mm} Y'_{3}=g(r),  \;\;
(dg/dr)^{2}=\frac{(r^{2}+r+1)}{r^{3}}\\
\vspace{1mm}
Y'_{4}=r\mbox{sin} \theta
\mbox{sin}\phi\\
\vspace{1mm}
Y'_{5}=r\mbox{sin}
 \theta \mbox{cos} \phi\\
\vspace{1mm}
Y'_{6}=r\mbox{cos}
\theta \vspace{1mm}\end{array}\right.\label{eq:YY}
\end{equation}
Notice that ${Y'}_{3}$ is defined  for  $r>0$, while $Y_{3}$ is defined only
for $r>1$, suggesting the  extension of  
$(E,g)$. In order to determine the metric $g'$
(extension of $g$), define the  new coordinates $u$  and $v$  by:\\
- For $r>2m$,
\begin{equation}
v=\frac{1}{4m}(\frac{r}{2m})^{1/2}exp(\frac{r}{4m}){Y'}_{1}\; \; \mbox{and}
\;\;u=\frac{1}{4m}(\frac{r}{2m})^{1/2}exp(\frac{r}{4m}){Y'}_{2}.
\end{equation}
- For $0<r<2m$,
\begin{equation} v=\frac{1}{4m}(\frac{-r}{2m})^{1/2}exp(\frac{r}{4m}){Y'}_{1}\;
\; \mbox{and} 
\;\;u=\frac{1}{4m}(\frac{-r}{2m})^{1/2}exp(\frac{r}{4m}){Y'}_{2},
\end{equation}
where
\begin{equation}
u^{2}-v^{2}=(\frac{r}{2m}-1)exp(\frac{r}{2m})\;\Longleftrightarrow \;{%
Y^{\prime }}_{2}^{2}-{Y^{\prime }}_{1}^{2}=16m^{2}(1-\frac{2m}{r}).
\end{equation}
Now $r=r({Y^{\prime }}_{1},{Y^{\prime }}_{2})$ is implicitly defined by last
equation, while $t=t({Y^{\prime }}_{1},{Y^{\prime }}_{2})$ is
implicitly defined by 
\begin{equation}
{Y^{\prime }}_{1}/{Y^{\prime }}_{2}=tgh(\frac{t}{4m}).\label{eq:tgh}
\end{equation}
Finally, the metric $g^{\prime }$ in the new coordinates results
\begin{equation}
ds^{2} = (32m^{3}/r) exp(-r/2m)(dv^{2} -
du^{2}) - r^{2}(d\theta ^{2} +
sin^{2}\theta d\phi ^{2}),
\end{equation}
Curiously this metric is exactly the same metric encountered by Kruskal. The
$u$ and $v$ coordinates,  $Q^2$ and all characteristics of Kruskal
metric are given by $(E'=Q^{2}\times S^{2},g')$, without a singularity at
$r=2m$. We know that $(E,g)$ is disconnected because it is composed by two
connected components. When we calculated the extension $(E',g')$ through the
Fronsdal  immersion we see that it is connected. \cite {EdMa}

In the following section, we will prove that the topology of $(E,g)$ is
different from   the topology of $(E',g')$.\cite{EdMa} 

\section {Schwarzschild's Brane-world Topology}

Let  $(U_{\alpha},\varphi _{\alpha})$ be  a  coordinate system 
on a point $p\in M^n$ of a differentiable manifold $M^n$. Generally speaking, the
topology of a manifold $M^n$ is defined naturally 
through its open sets.  If
$A\subset M^n$,  then $A$ is an open set of $M^n$ if 
$\varphi _{\alpha}(A\cap
\varphi _{\alpha}^{-1}(U_{\alpha}))$ is an open set of $\Re^n$, $\forall \alpha$.
In other words,  the atlas of $M^n$ determines its topology.\cite{Oneill}

The following theorem  shows that
the topology of $(E,g)$ is different from that of $(E',g')$.
%\begin{theorem}
\vspace{3mm}

\underline{\bf Theorem:}
%\vspace{3mm}

{\it The topology of a
gravitational field outside of a body with spherical symmetry is given by
$\Re^{2}\times S^{2}$.}
%\end{theorem}
%\begin{proof}

\underline{\bf Proof:} 
\vspace{3mm}

By  construction,  $E=[P_{I}^{2}\cup P_{II}^{2}]\times S^{2}$ and $E'=Q^{2}\times S^{2}$.
The topology of $E$ is the Cartesian product topology of $[P_{I}^{2}\cup
P_{II}^{2}]$ by $S^{2}$, while that the topology of $E'$ is the Cartesian
product topology of $Q^{2}$ by $S^2$. The topology of $S^{2}\subset \Re^{3}$
is the usual  topology induced by  the topological space $(\tau
_{3},\Re^{3})$. On the other hand, the topologies of
$[P_{I}^{2}\cup P_{II}^{2}] \subset \Re^{2}$ and of $Q^2 \subset \Re^{2}$,
respectively  $\tau_{p}$ and $\tau _{q}$,  will be induced from $(\tau
_{2},\Re^{2})$. 
Since  $Q^{2}$ is an extension of $[P_{I}^{2}\cup
P_{II}^{2}]$,  we may define one isometric immersion, 
\[
\psi :\;[P_{I}^{2}\cup
P_{II}^{2}]\;\longrightarrow\;Q^{2}.
\]
Therefore, for an  open set $A\subset \Re^{2}$ given by
\[
A=\{(t,r)\in \Re^{2}| \;t^{2}+(r-2m)^{2}<m^{2}\;\,\mbox{and}\;\, r>0\}.
\]
we have 
$A\cap [P_{I}^{2}\cup P_{II}^{2}]=A-\{(t,r)\in \Re^{2} | \;r=2m\}$. This 
is an open set of the topological space  $[P_{I}^{2}\cup P_{II}^{2}]$,
composed of two connected components. Observe that open sets form a
topological basis for the semi-plane $ t-r,\;r>0$. However, we have
that 
$\psi(A\cap [P_{I}^{2}\cup P_{II}^{2}])$ is
given for an open set composed by four connected components. As the lines
$L_{1}$ and $L_{2}$ defined for $r=2m$ from equation $(2)$ are on $Q^{2}$
we have that 
\[
\{[\psi (A\cap
[P_{I}^{2}\cup P_{II}^{2}])\cup L_{1}\cup L_{2}]\}\cap D = B ,
\]
where $D$
is an open disk on $\Re^{2}$ with center in the origin of $Q^{2}$. The set 
$B$ is a  plane disk, in the new coordinates $r=r({Y'}_{1},{Y'}_{2})$
and $t=t({Y'}_{1},{Y'}_{2})$.
In this manner the topology
of $Q^2$ is given by open sets of $\Re^2$. Finally 
we have that the topology of
$(E,g)$ is equal to $(\Re^{2}-\{(t,r)\in \Re^{2} |\;r=2m\})\times S^{2}$, clearly different
of the topology of space-time $(E',g')$ that is $\Re^{2}\times S^{2}$.$\triangle$\cite{EdMa}
%\end{proof}

Next section we investigate changes in the bulk's signature when we have  
changes in the topology of brane-world, under some conditions.

\section{Change in the Bulk's Signature}
%\begin{theorem}
%\label{theo:ff}

{\bf\underline{Theorem}:}
%\vspace{3mm}

{\it Let $(M^{n},g)$ and $(\bar{M}^{D},\bar{g})$, $D\geq n$, pseudo-Riemannian
manifolds and consider $Y:(M^{n},g)\longrightarrow
(\bar{M}^{D},\bar{g})$ a local isometric immersion. Let
$\tau_{\eta}$ the topology of the $Y(W^{n})$ isometric immersed submanifold of 
$(\bar{M}^{D},\bar{g})$, $W^{n}$ a neighborhood of $p \in (M^{n},g)$. If we change
topology of $Y(W^{n})$ to $\tau'_{\eta}$, and if $det(g_{ij})$, 
$det(g'_{\mu \nu})$ differ in sign at a point, then there exist a
change of signature of form assigned to $\bar{g}$.}
%\end{theorem}
%\begin{proof}

{\bf\underline{Proof}:} 
%\vspace{3mm}

This result basically came from linear algebra and from
differentiable manifolds elementary theory. Suppose that
$\eta_{1}$, $\eta_{2}$ are charts of $Y(W^{n})$ with intersecting
domains $U, V$. Then $\eta_{2}\circ\eta_{1}^{-1}$ is a
diffeomorphism and so its domain $\eta_{1}(U\cap V)$ must be open
in $\Re^{D}$. Since $(U\cap V)$ is a subset of $U$ such that
$\eta_{1}(U\cap V)$ is open in $\Re^{D}$, then
$\eta_{1}\mid_{(U\cap V)}$ is a chart of $Y(W^{n})$ with domain
$(U\cap V)$. These arguments are sufficient to conclude that the
collection of coordinate domains of manifold $Y(W^{n})$ forms a
basis for a topology on the set $Y(W^{n})$. The topology thus
induced on the set $Y(W^{n})$ by its $C^{\infty}$ structure is
called $\tau_{\eta}$ topology of the manifold $Y(W^{n})$. With
this topology, a non-empty subset $U$ of $Y(W^{n})$ is open iff
each point of $U$ has a coordinate neighborhood which lies in $U$.
\cite{Bri}

We note that if $Y:(M^{n},g)\longrightarrow (\bar{M}^{D},\bar{g})$ is a local
isometric immersion, then $\exists p\in W_{n}\subset M_{n}$, $W_{n}$ a
neighborhood of $p$, such that $Y\mid _{W_{n}}$ is an embedding, and if
$Y(W_{n})$ is assigned the metric tensor such that the induced map
$W_{n}\longrightarrow Y(W_{n})$ is an isometry, then $Y(W_{n},g)$
is a pseudo-Riemannian submanifold of $(\bar{M}^{D},\bar{g})$. In other words
we say that locally an isometric immersion is essentially a
pseudo-Riemannian submanifold.

Denote  $\tau_{\eta}$ the topology of $Y(W^{n},g)$ assigned to
atlas $A$ and another topology $\tau'_{\eta}$ of $Y(W^{n},g)$
assigned to new atlas $A'$, both induced topologies on the set
$Y(W^{n},g)$ by its $C^{\infty}$ structure.

Let $g_{ij}$ the components of metric tensor $g$ on the differential
quadratic form, $g_{ij}dx^{i}dx^{j}$. By a real transformation
this quadratic form at a point $p\in W^{n}$ is represented by
\[
(dx^{1})^{2}+...+(dx^{r})^{2}-(dx^{r+1})^{2}-...-(dx^{n})^{2},
\]
where $S=2r-n$ is the signature of the form.

Suppose it exists a transformation such that
\[
g'_{\mu\nu}=g_{ij}\frac{\partial x^{i}}{\partial
x'^\mu}\frac{\partial x^{j}}{\partial x'^\nu},
\]
from the rule for multiplication of determinants we have
\[
det(g'_{\mu\nu})=det(g_{ij})J^{2},
\]
where $J$ is the Jacobian of the transformation from $g_{ij}$ to
$g'_{\mu\nu}$. Now suppose that $det(g_{ij})$ and
$det(g'_{\mu\nu})$ differ in sign at a point, for instance at $p\in
(W^{n},g)$. Thus we have an imaginary transformation from $g_{ij}$
to $g'_{\mu\nu}$ at $p\in (W^{n},g)$. \cite{Eis}\\ 
For this transformation at $p$ the form $g_{\mu\nu}dx'^{\mu}dx'_{\nu}$ can
be represented by
\[
(dx'^{1})^{2}+...+(dx'^{r'})^{2}-(dx'^{r'+1})^{2}-...-(dx'^{n})^{2},
\]
where $S'=2r'-n$ is the signature of this form, clearly $S\neq
S'$.

We note that $g$ is the induced metric of $\bar{g}$, thus we have
 $g(u,v)=\bar{g}(dY(u),dY(v)),\forall u,v \in T_{p}(W^{n},g)$. 
Using the facts before it is easy to see that
exists a change of signature of form assigned to $\bar{g}$ at
point $Y(p) \in Y(W^{n},g)$. We remember that for real transformations the signature
of $\bar{g}$ is invariant.$\triangle$
%\end{proof}

\section{The Schwarzschild's Brane-world Immersed into $(\bar M^{6},\bar g)$}

Randall and Sundrum have proposed an interesting scenario of extra
non-compact dimensions in which four-dimensional gravity emerges
as a low energy effective theory, to solve the hierarchy problems
of the fundamental interactions.\cite{RS}This proposal is based on the
assumption that ordinary matter and its gauge interactions are
confined within a four dimensional hypersurface, the physical
brane, immersed in a five-dimensional space of constant curvature. In order to describe
the real world, the Randall-Sundrum scenario has to satisfy all
the existing tests of General Relativity, with base in the motion
of material particles within the Schwarzschild's brane-world that is
given by four-dimensional geodesic equation. That is an excellent
and famous model, but from the point of view of mathematics and
of cosmological observations some constraints exist. The principal problem that appears on the mathematical 
model of the brane-worlds has origin on the choose of bulk, when we require a compatible immersion of 
the Schwarzschild's brane-world into five-dimensional bulk of constant curvature.\cite{Monte,Ruth} An alternative bulk to 
Randall-Sundrum model is when we consider the case of six-dimensional 
flat bulk, where the Schwarzschild's brane-world has a compatible immersion in 
$(\bar M^{6},\bar g)$.\cite{monte} 

According to section (2) to (5) we concluded that brane-world $(E,g)$ is disconnected because it is composed by two
connected components. We have that the topology of  $(E,g)$ is given by, 
$(R^{2}-\{(t,r)\in \Re^{2} |\;r=2m\})\times S^{2}$ that is different
of the topology of the brane-world $(E',g')$ which is equal to $\Re^{2}\times
S^{2}$, this latter is due to the theorem (1). 
For a {\it trivial example of the theorem (2)}, suppose that space-time $(E,g)$ is immersed
into a pseudo-Euclidean manifold of six
dimensions $(\bar M^{6},\bar g)$, where $\bar g$ is given by Kasner and {\it signature} $S=-2$, as we can see from section (2). 
Now from the new coordinates, suppose that $(E',g')$ is
immersed into a pseudo-Euclidean manifold of
six dimensions, $(\bar M^{6},\bar g)$, we must have {\it a change of signature} from  $S=-2$ to $S'=-4$, for the
$\bar g$ metric that is given by Fronsdal, (section (2)).

\section{Comments}
We prove that beginning from a brane-world immersed in
an environment space of higher dimensional (bulk) it is possible to
change the bulk's signature when we change the topology of brane-world. This
result show us that we can have nature geometric and topological
constraints which can to enjoin some link with physical
properties of the space-time on brane-world context. For instance, the fact that we
assume the brane-world $(E',g')$ to be connected, because
disconnected components of the universe cannot interact by means
of any signals and the observations are confined to the connected
component wherein the observer is situated. In this case, some constraints appear in the bulk's signature $(\bar M^{6},\bar g)$, 
where we can to notice that $\bar g$ metric must have only one time coordinate.

\section{Acknowledgments}
The author would like to thank Professor Marcos D. Maia for useful
discussions, to FAPES-ES-CNPq-PRONEX-BRASIL and to CAPES-BRASIL (grant- 3749/09-6-senior stage) for partial financial support.


\begin{thebibliography}{20}

\bibitem{MonteIJGMMP} Edmundo M. Monte, Mathematical support to brane-world theory, {\it Int. J. Geom. Meth. Mod. Phys} {\bf 4} (2008) 1259-1267.

\bibitem{Eis}  L. P. Einsenhart, {\it Riemannian Geometry} (Princeton University Press, New Jersey, 1966).

\bibitem {Kanti} P. Kanti, et al, Schwarzschild black branes and strings in higher-dimensional brane-worlds, hep-th/0207283.

\bibitem{HE}  S. Hawking and G. Ellis, {\it The Large Scale Structure of Space-Time} (Cambridge University Press, Cambridge, 1973).

\bibitem{Oneill}  B. O'Neill {\it Semi-Riemannian Geometry} (Academic Press, New York, 1970).

\bibitem{Kasner} E. Kasner, Finite representation of the solar gravitational field in flat space of six dimensions, {\it Am. J . Math.} {\bf 43} (1965) 130-133.

\bibitem{Fronsdal} C. Fronsdal, Completion and embedding of the Schwarzschild solution, {\it Phys. Rev.} {\bf 116} (1959) 778-781. 

\bibitem{Kruskal}  M. Kruskal, Maximal extension of Schwarzschild metric, {\it Phys. Rev.} {\bf 119} (1960) 1743-1745.

\bibitem {EdMa} Edmundo M. Monte and M. D. Maia, Topology and extension of Schwarzschild, {\it Mat. Cont.} {\bf 13} (1997) 229-233.

\bibitem{Bri} F. Brickell and R. Clark, {\it Differentiable Manifolds} (Company London, N.York, 1970).

\bibitem{RS} L. Randall and R. Sundrum, A large mass hierarchy from a small extra dimension, {\it Phys. Rev. Lett.} {\bf 83} (1999) 3370-3373.

\bibitem{Monte} Edmundo M. Monte, The embedding of Schwarzschild in braneworld, {\it Int. J. Theo. Phys.} {\bf 48} (2009) 409-413.

\bibitem {Ruth} R. Durrer, et al, Testing brane worlds with the binary pulsar, hep-th/0305181.

\bibitem {monte} M. D. Maia and Edmundo M. Monte, On the Stability of Black Holes at the LHC, {\it Spanish Relativity Meet. ERE2010}, Granada, Spanish, Sept. 2010 and hep-th/08082631.
  
\end{thebibliography}
\end{document}